\documentclass[12pt,epsf]{article}

\begin{document}

\begin{flushright}
SLAC-PUB-9620\\
January 2003\\
\end{flushright}

\bigskip\bigskip
\begin{center}
{\bf\large On E.D.Jones'
MICROCOSMOLOGY\footnote{\baselineskip=12pt Work supported by
Department of Energy contract DE--AC03--76SF00515.}}
\end{center}

\begin{center}
H. Pierre Noyes, Stanford Linear Accelerator Center MS 81,
Stanford University, 2575 Sand Hill Road, Menlo Park CA 94025,
noyes@slac.stanford.edu\\
Louis H. Kauffman, Department of Mathematics, University of
Illinois at Chicago, 851 South Morgan Street, Chicago, IL 60607-7045, kauffman@uic.edu\\
James V. Lindesay, Physics Department, Howard University,\\
Washington, D.C. 20059, jlslac@slac.stanford.edu\\
Walter R. Lamb, Energy Systems - Solar Inc.\\
148 Jacinto Way, Sunnyvale, CA 94086, LWrlamb@cs.com\\
\end{center}

\begin{center}
{\bf Abstract}
\end{center}

By taking seriously the limits on observability which come from
combining relativistic quantum mechanics with general relativity,
Ed Jones has shown that the current measurements of the
cosmological constant density $\Omega_{\Lambda}\sim 0.7$ imply
that the temperature scale at which it becomes possible to discuss
cosmological models is $\sim 5 \ Tev$ ($5.8\times 10^{16} \ ^oK$).
This is self-consistent with the assumption that the number of
Planck masses which make some sort of ``phase transition'' to this
state is $N_{Pk}\sim 4\times 10^{61}$. We review Jones' argument
and the {\it bit-string physics} calculation which gives the
baryon-photon ratio at nucleosynthesis as $\sim 2/256^4$, the dark
matter-baryon ratio as $\sim 12.7$, and hence $\Omega_m \sim 0.3$,
in agreement with current observations. Accepting these values for
the two energy densities $\Omega_{\Lambda}+\Omega_m \sim 1$ in
accord with recent analyses of fluctuations in the CMB showing
that space is flat to about 6\%. We conclude that experiments with
particle accelerators in the 5-10 Tev range must either show that
current theory can adequately describe the currently observed
structure of our universe or force us to revise our ideas about
physics at a very fundamental level.

\bigskip

\begin{center}
Presented at The Twenty-Fourth Annual Meeting of the\\
{\bf ALTERNATIVE NATURAL PHILOSOPHY ASSOCIATION}\\
Cambridge, England, 15-20 August 2002\\
\end{center}

\bigskip

\begin{center}
{\bf I. INTRODUCTION}
\end{center}

It has been known for some time that standard general relativistic
cosmology can describe current astronomical observations in
remarkable detail.\cite{PDG00,PDG02} Retrodicting from these
results to earlier times, one finds extremely hot and dense
radiation and matter $\sim 13 \ Gyr$ ago. Conditions were such
that no current experimental or astronomical systems are available
to test what the ``laws of physics" were then, or even whether
such laws ``existed''. As E.D.Jones realized some time ago, one
way to make progress in gaining a limited understanding of this
``pre-physics" era is to take seriously the limitations that the
shortest measurable length and time, and the largest measurable
elementary mass and temperature place on physical cosmology. In
this paper we review and discuss the remarkable discoveries that
he made. We emphasize their fundamental character and independence
of special assumptions.

E.D.Jones\cite{Jones97} discovered that a {\it neo-operational}
\cite{Noyes96} approach imposes fundamental limits on the
measurement of short distances in such a way that we can {\it
predict} that there must be a {\it positive} cosmological
constant. Further, his preliminary calculation gave the
cosmological constant energy density, relative to the critical
density, of $\Omega_{\Lambda}= 0.6\pm 0.1$. The basic idea is that
we must find some way to connect the inchoate, pre-geometrical,
pre-physical state of the universe by a ``phase transition" to a
state in which contemporary physics --- explicitly, relativistic
quantum mechanics and general relativity
--- can be consistently employed. The connection will be made by
using one scaling parameter. This scaling parameter is fixed by
requiring energy density equilibrium between the virtual energy
density which thermalizes at some mass scale $m$  and the virtual
energy density $\rho_{\epsilon} ={3\epsilon\over 4\pi
r_{\epsilon}^3}$ which is ``left behind" when the (extremely
rapid) transition to ordinary space, time and particles is
complete. Here $r_{\epsilon}= \hbar c/\epsilon$. Since this
residual virtual energy  ${\epsilon}$ necessarily decouples from
other contributions to the FRW equations for the evolution of the
universe, Jones' interpretative postulate is to identify it with
the cosmological constant density, i.e.
$\rho_{\epsilon}=\rho_{\Lambda}$.

The limiting concepts we apply to obtain limits on observability
are the uncertainty principle, the black hole surface area, and
the cosmological event horizon. We make contact with experimental
practice by assuming that we know four basic universal constants
in standard units. These are:

(1) Newton's gravitational constant ${\bf G_N}$, which can be
fixed if we know the escape velocity of a particle from a system
of radius $R$, and mass $m$, i.e.
\begin{equation}
{{\bf G_N }m \over R} ={1\over 2}v^2_{escape}
\end{equation}

(2) Boltzman's constant ${\bf k}$ from the basic law of
statistical mechanics connecting entropy, $S$, to the number of
degrees of freedom $W$, i.e.
\begin{equation}
S= {\bf k} \ ln  \ W
\end{equation}

(3) Planck's constant ${\bf h}$ (or $\hbar=h/2\pi$), as in
\begin{equation}
E={\bf h} \ \nu
\end{equation}

(4) Einstein's limiting velocity ${\bf c}$, as in the mass-energy
relation
\begin{equation}
E=m \ {\bf c}^2
\end{equation}

\begin{center}
{\bf II. Shortest length; largest density; highest temperature}\\
\end{center}

If there is a largest {\it elementary} fermion mass $m_X$, the
shortest geometrical length in (special) relativistic quantum
mechanics to which we can examine its (or any) structure is the
Compton radius
\begin{equation}
R_Q (m_X)={\hbar\over 2m_X c}
\end{equation}
Wick pointed out long ago\cite{Wick38} that if we try to measure
the structure of a system which couples to some mass $\mu$,
combining special relativity with the quantum energy-time
uncertainty principle tells us that at distances less than
$\hbar/\mu c$, which requires us to use available energies greater
than $\mu c^2$, such a particle can be produced with a finite
probability. For particles of mass $m_X$ that satisfy the CPT
theorem --- no exceptions are known --- and have a conserved
quantum number which allows anti-particles to be defined (which
includes all known fermions), examining the structure of $m_X$
with {\it any} probe of energy greater than $2m_Xc^2$ will
necessarily produce a pair consisting of $m_X$ and its
anti-particle with some finite probability. But the $m_X$ produced
will be indistinguishable from the $m_X$ probed, and can appear
{\it anywhere} within a distance $\hbar/2m_X c$ of the target
particle. Thus, in relativistic quantum mechanics (in contrast to
non-relativistic quantum mechanics which in principle allows any
short distance to be examined by using probes of high enough
momenta) there is an absolute shortest distance, set by the
largest elementary mass. Below this limit the concepts of point
geometry simply {\it dissolve}.

In contrast, general relativity {\it starts} with geometrical
concepts and would seem to be able (until we come to cosmology) to
describe systems with arbitrarily large masses. However, if the
mass is concentrated within a volume whose surface area is (for
spherical symmetry) $4\pi R_G(m)^2$, we must use more care.
``Inside" the {\it black hole} defined by this area and calculable
from the Schwarzschild radius
\begin{equation}
R_G(m) = {2mG_N\over c^2}
\end{equation}
where $G_N$ is Newton's gravitational constant, there is no known
way to {\it measure} geometrical structure. Only the mass, and (if
it has them) charge and magnetic moment can be measured by means
of {\it exterior} observations. Consequently, in regions where the
mass concentrations are small enough, we can use (if available)
systems with small enough mass-energies to probe geometrical
structure down to arbitrarily short lengths. Thus, by itself,
general relativity seems to be able to use geometrical concepts
consistently in any regions {\it outside} of black holes.

But any theory of {\it quantum gravity} must, at least in a
limiting sense, combine both relativistic quantum mechanics and
general relativity in a single theory. Since the shortest length
observable in the first theory is inversely proportional to the
mass considered, while the shortest length observable in the
second theory is proportional to the mass considered, this
requirement tells us (in the absence of new concepts) that there
must be a finite limiting length for {\it any} theory of quantum
gravity. We define this by setting the Schwarzschild radius equal
to the Compton radius, i.e.
\begin{equation}
R_{min}^{QG}=R_G(m)=R_Q(m)
\end{equation}
This leads immediately, to the requirement that $4m^2 = \hbar
c/G_N \equiv M_{P}^2$.

The limiting mass $M_{P}$ calculated in this way was first arrived
at by Planck (using  $h$), and is now (using $\hbar$) called the
{\it Planck mass}. Planck's original argument was simply that
$G_N$, $k$, $c$, and $h$ were clearly universal constants. Since
their dimensions in terms of (sometimes fractional) powers of
mass, length, time, and temperature are {\it independent}, these
constants collectively define what are certainly universal, and
perhaps fundamental dimensional units for physics. But the
numerical values so predicted were discouraging. The {\it Planck
mass} in SI units is approximately $2.18\times 10^{-8}$ kilograms.
Even today there is no known elementary particle with this mass.
The corresponding {\it Planck length} $L_{P}\equiv
\hbar/M_{P}c=1.62\times 10^{-35} \ m$, {\it Planck time} $T_{P}
\equiv \hbar/M_{P}c = 5.39\times 10^{-44} \ sec$ and {\it Planck
temperature} $\Theta_{P} \equiv M_{P}c^2/k = 1.42\times 10^{32} \
^oK$. These sounded ridiculous if one wished to relate them to
available methods of measuring masses, lengths, times and
temperatures. But as {\it limiting} quantities, they make sense.
To our knowledge, Jones' simple calculation, which at least for us
makes this interpretation compelling, originated with him.

That it makes no sense using current theory to talk of lengths
less than the Planck length (or times less than the Planck time,
or elementary masses greater than the Planck mass, or temperatures
higher than the Planck temperature) is hardly a novel conclusion.
But Ed Jones' demonstration of the fact, which we have just
summarized, is so straightforward that we think it deserves wide
circulation.

What is, perhaps, less familiar is that the same physical
principles prohibit meaningful discussion of physical matter at
any density greater than the Planck density, i.e. a Planck mass in
a Planck volume which must exist for at least a Planck time, that
is for at least $[{\hbar G_N\over c^5}]^{{1\over 2}}$. We define
this limiting  density scale by
\begin{equation}
\rho_P \equiv {M_P\over {4\pi \over 3}  L_P^3}= {3\over
4\pi}[c/\hbar]^3M_P^4
\end{equation}
From now on we will use units in which $\hbar=1=c=1=k=1$, so that
any mass $\mu$ can stand (ambiguously) for either a mass or an
energy or a temperature; any inverse mass $\mu^{-1}$ can stand
(ambiguously) for a length or a time. We see that in these units
${3\over 4\pi}\mu^4$ is the energy density of mass-energy $\mu
c^2$ in a spherical volume whose radius is its Compton wavelength.

We now show that in these units ${3\over 4\pi}M_P^4$ is, indeed,
the limiting energy density. Consider first masses less than a
Planck mass inside a Planck volume. For such systems, the Compton
wavelength exceeds the Planck length, and the limiting density is
achieved only when the contained mass is equal to the Planck mass.
Consider next systems with a mass greater than the Planck mass
which we try to confine within a Planck volume. For such systems,
the Schwarzschild radial coordinate defined by the horizon area
exceeds the size of the system we are attempting to construct, and
to assume that it is confined within a smaller radial coordinate
is, again, meaningless. Although the radial coordinate does not
measure proper distance, it nonetheless sets the radial scale of
the geometry relative to the horizon. So only the limiting case of
systems at the Planck density (or less) whose volume just exceeds
the volume measured by the Schwarzschild radius (or is greater)
have any hope of being given observational meaning. Since these
two cases exhaust the possibilities, no system with greater than
the Planck density can be given operational meaning. Q.E.D.

For completeness, we also note that the same considerations we
used to establish limits on elementary masses, lengths and times
also limit the maximum temperature to the Planck temperature. At
first sight, this is not completely obvious. We are allowed
massless radiation, and hence could define temperature by the
Planck distribution in, for example, an Einstein-De Sitter
universe that contains only gravitation and electromagnetic
radiation. However radiation, by itself, cannot come to
equilibrium even when the individual frequency modes are quantized
using the $E=h\nu$ prescription for each degree of freedom. These
modes could be initially set to any arbitrary distribution.
Without some interaction between the modes this arbitrary initial
distribution will never come to the Planck distribution. Hence,
there is no way to define ``temperature".

To obtain insight on how relativistic quantum mechanical
temperature measurement necessarily gets connected to a mass scale
it is useful to go back to the classic paper by Bohr and
Rosenfeld\cite{Bohr&Rosenfeld33} which provided the initial
operational justification for ``second quantization" of the
electromagnetic field.  They derived the commutation relations
between electric and magnetic field strengths by applying the
uncertainty principle to the material apparatus which measures
them. This derivation can avoid any discussion of the particulate
structure of matter because the only universal constants they
invoke are $\hbar$ and $c$. Thus the theory they discuss is scale
invariant and their analysis need only use wavelengths so long
that the classical description of matter can be invoked
consistently. However, as they point put, when wavelengths of the
order of $\hbar/mc$ are required, which is true at mass scale $m$,
their derivation breaks down. We then need to use the coupling
between the matter field and the electromagnetic field in order to
study the meaning of measurement. This, in turn, brings in the
scattering of light by light due to the creation (real or virtual)
of particle-antiparticle pairs. This coupling then provides
transitions between electromagnetic modes and in the black body
situation leads, sooner or later, to the Planck distribution.
Clearly this same mechanism can achieve thermalization in the
cosmological context. Once more this ties us to some finite mass
scale, and the arguments already given limit us to temperatures
below the Planck temperature.

Since any system we consider, including our observed universe,
cannot be described using current physics if it exceeds the Planck
density, we next ask whether assigning the Planck density to our
universe at some stage in its history makes sense. Consider first
the case when the universe that starts at this density has one
Planck mass. But then its event horizon, or radial scale factor in
the FRW metric, coincides with the Schwarzschild radius for that
mass. There is no space ``outside" this radius which can allow us
to define the volume which this mass occupies, so the concept of
density remains vacuous. However, if this universe has a larger
radial parameter (and hence a lower density) allowing it to become
a ``black hole'' surrounded by an ``ergo-sphere'' made up of
particles and anti-particles  and radiation at some
mass-energy-temperature scale $m < M_P$, it becomes a possible
starting point for a universe which becomes describable at an
event horizon parameterized by $R_H \sim 1/m$. Jones realized that
these arguments are closely related to the Dyson-Noyes
argument\cite{Noyes75,Noyes97}, as we now show.

Dyson\cite{Dyson52} pointed out that if there are
$Z_{e^2}=\alpha_{e^2}^{-1}\simeq 137$ electromagnetic interactions
within the Compton wavelength of a single charged
particle-antiparticle pair (i.e. $\hbar/2mc$), there is enough
energy to create another pair. Whether these interactions are
virtual, or real (eg in a system with enough energy and
appropriate internal momenta to concentrate $2mc^2Z_{e^2}$ of that
energy within this Compton wavelength), in a theory for which like
charges attract rather than repel each other still more energy can
then be gained by creating another pair; the system collapses to
negatively infinite energy. Dyson concluded that the renormalized
perturbation theory for QED is not uniformly convergent beyond 137
terms. Note that this bound can be written as $Z_{e^2}\alpha_{e^2}
= 1$. Noyes\cite{Noyes75,Noyes97} noted that for electron-positron
pairs, this critical energy corresponds approximately to the
threshold for producing a pion. This fact provides a physical
interpretation of the reason for the failure of QED: QED ignores
strong interactions mediated by pions, or more generally by quarks
and anti-quarks which bind to yield pions.

For gravitation the corresponding coupling constant $\alpha_m
=G_Nm^2 =m^2/M_P^2$ and the critical condition becomes
\begin{equation}
Z_m\alpha_m = 1 \ or \ Z_m = {M_P^2\over m^2}
\end{equation}
where $Z_m$ represents the number of gravitational interactions
within $\hbar/mc$ defining this critical condition. That is, for
quantum gravitational perturbation theory, the cutoff mass-energy
corresponds to the Planck mass rather than the pion mass, which
makes sense.

\begin{center}
{\bf III. Plancktons: The Pre-physics - Physics Transition}\\
\end{center}

To see how a dense system like that discussed at the end of the
last section  might make sense in a cosmological context, consider
first a system which starts from a Planck's mass worth of
quantized mass-energy (at some mass scale $m$ where we have
confidence that currently accepted physics is applicable)
distributed in a radial shell with radially inward-directed
momentum. Such a system, assuming no changes in the physics along
the way, would have a finite chance of making it down to the
Planck density scale before rebounding (or whatever). This
suggests that we might, in some sense, be able to to describe a
universe which ``starts" with one {\it Planckton} (which we define
as a Planck mass in a Planck volume and hence necessarily at the
Planck temperature). If our current physics is not capable of
describing a single Planckton, it is still possible to envisage
some non-adiabatic expansion process which would allow the virtual
energy ``contained" in this Planckton to be distributed throughout
a large volume at some small enough mass scale and event horizon
large enough so that {\it after this expansion} this universe {\it
will be describable} using currently understood physics. This
assumes, of course, that the mass scale $m$ is low enough that we
have confidence that current theory can describe $Z_m$
gravitational interactions within volumes contained within the
event horizon whose linear measure is $\sim 1/m$, and that the
Dyson-Noyes analysis applies.

Of course the universe described in the last paragraph is not {\it
our} universe, which is known to contain (and have contained for
twelve thousand million years or more) a lot more than one
Planckton's worth of mass. However, since we are postulating a
non-adiabatic transition from the Planck density to situations we
{\it can} describe, it makes just as much sense to start from
$N_{Pk}$ Plancktons as to start from one. Of course this whole
assemblage would have to be at the Planck density or just below
it; on another occasion we hope to be able to discuss whether such
an assemlage can be consistently described as a state of {\it
Plancktonic matter}. This idea allows us to think of the initial
expansion as starting from a virtual energy state and making a
transition to a real energy state at some mass scale $m$, {\it
leaving some of that virtual energy behind}.

Assume that the event horizon at which this residual virtual
energy $\epsilon$ has observable consequences can be characterized
by a radial parameter which we call $R_H(\epsilon)\sim
1/\epsilon$. Further assume that the transition point corresponds
to some expansion factor $Z_{\epsilon}$ from the Planck length,
i.e.
\begin{equation}
R_H(\epsilon) \equiv Z_{\epsilon}L_P={Z_{\epsilon}\over M_P} \sim
{1\over \epsilon}
\end{equation}
Here $\epsilon$ is the energy {\it per Planckton}. One might note
that this defines the expansion factor in terms of $R_H$, which
will give a measure of the square root of the area of the horizon,
NOT a radial distance. This is a key feature of holographic
interpretations of the physical properties of
horizons\cite{Bigatti&Susskind00}. This relationship also
associates the energy per Planckton with the expansion scale of
this horizon from the Planck length.

The corresponding energy density scale which this residual virtual
energy has at the moment when the phase transition is complete is
$\rho_{\epsilon} = N_{Pk}{3\over 4\pi}\epsilon^4$. Since this came
from $N_{Pk}$ Plancktons, each confined within a volume scale
given by a single Planckton, we can also write down the event
horizon (scale factor, ``Schwarzschild radius of the universe")
$R_H(\epsilon)$ defined by this mass scale as
\begin{equation}
R_H(\epsilon)=G_NN_{Pk}\epsilon \Rightarrow 1 \sim
{N_{Pk}\epsilon^2\over M_P^2}= {N_{Pk}M_p^2\over
M_P^2Z_{\epsilon}^2} \Rightarrow N_{Pk}=Z_{\epsilon}^2
\end{equation}

Jones now assumes that the bulk of the energy which started out as
$N_{Pk}$ Plancktons thermalizes at mass scale $m$ and that the
corresponding energy density $\rho_m ={3\over 4\pi}m^4$ is in
energy equilibrium with the residual energy $\rho_{\epsilon}$ at
the moment of completion of the phase transition. Hence
\begin{equation}
N_{Pk}\epsilon^4=m^4 \Rightarrow {N_{Pk}\over Z_\epsilon^4}
={m^4\over M_P^4}={1 \over Z_m^2} \Rightarrow Z_m^2=Z_{\epsilon}^2
\end{equation}
Thus the Dyson-Noyes factor $Z_m$ (which allows for operationally
definable coordinates in the quantum domain) coincides with the
residual virtual energy expansion scale factor $Z_{\epsilon}$.
This is true only at the completion of the non-adiabatic phase
transition (at which point there are also operationally definable
coordinates in the gravitational/geometrical domain). Hence we
conclude that
\begin{equation}
Z_{\epsilon}=Z_m=Z=N_{Pk}^{{1\over 2}} \ \ \ and \ \ \
m^2=\epsilon M_P
\end{equation}

Jones also points out that this fact is over-constrained because
$Z_{\epsilon}$ and $Z_m$ each corresponds to the entropy (number
of degrees of freedom) of each of the systems, which must also be
equal at equilibrium. We also note that the mass scale at which
the transition becomes complete is the geometric mean between the
residual virtual energy and the Planck mass. Since the system
starts in the pre-physics regime in which geometric structure
cannot be specified, in the resulting thermalized state, prior to
further evolution, ``where" in the earlier state any Planckton's
worth of energy ``came from" also cannot be specified. Hence the
virtual energy which is momentarily in density equilibrium with
the mass-energy at mass scale $m$ is {\it uniformly} distributed.
Consequently the universe describable using current physics {\it
starts out} with no structure even though it is much too large for
the regions to be causally connected. Thus, Jones' use of
fundamental physical principles solves the ``horizon problem" {\it
without} having to postulate the unknown ``physics" implicit in
the currently popular ``inflationary" scenarios.

We now complete the Jones argument. Note that since the residual
energy density must be positive, and since the transition
--- whatever the details --- must be extremely rapid, it can be
seen that $\rho_{\epsilon}$ corresponds to the cosmological
constant density ``boundary condition" $\rho_{\Lambda}$ in the FRW
equations. That it is positive is required for logical
consistency, because this corresponds to a ``negative pressure"
(expansive force)\cite{Frieman02a} which makes the transition {\it
irreversible}.

We can now check this conclusion by comparison with cosmological
observations. Putting together the algebraic results already
established and the definition of the cosmological constant energy
density in terms of the critical density
($\rho_{\Lambda}=\Omega_{\Lambda}\rho_c=\Omega_{\Lambda}\rho_{\epsilon}$),
we have that the basic scale parameter
\begin{equation}
Z=({\rho_P\over \Omega_{\Lambda}\rho_c})^{{1\over 4}}=({0.7\over
\Omega_{\Lambda}})^{{1\over 4}}({0.71\over h_0})^{{1\over
2}}\times 6.564\times 10^{30}
\end{equation}
For the Planck density we use Eq. 8, which works out to be
$\rho_P=6.906\times 10^{117} {Gev/c^2\over cm^3}$. For the
critical energy density we use\cite{PDG00}
$\rho_c=1.054\times10^{-5}h_0^2 \ Gev/c^2 \ cm^{-3}$. Finally we
accept $\Omega_{\Lambda}=0.7$ for the normalized cosmological
constant density and $h_0=0.71$ for the normalized Hubble
constant, as is indicated in Eq. 14. It is then easy to calculate
the thermalization mass scale from the Dyson-Noyes relation (Eq.
9) as
\begin{equation}
m=Z^{-{1\over 2}}M_P=({ \Omega_{\Lambda}\over 0.7})^{{1\over
8}}({h_0\over 0.71})^{{1\over 4}}\times 4.766 \ Tev/c^2
\end{equation}

The positive cosmological constant, let alone its value, was still
a matter of debate two or three years ago, resting as it did
solely on the measured luminosity and red shifts of a number of
distant type Ia supernovae (IaSne). These results have recently
been improved\cite{PDG02}. As was pointed out this spring by
Frieman\cite{Frieman02b}, lingering doubts can be set to rest by
the fact that a completely different type of evidence now shows
that $\Omega_{\Lambda}=0.7$, consistent with the Type Ia supernova
data. The new evidence is simply that {\it fluctuations} in the
cosmic microwave background radiation show that our universe is
{\it flat} to about 6\%, i.e. that $\Omega_m +\Omega_{\Lambda} =
1$ where $\Omega_m$ is the normalized mass-energy density. Since
it is known from a number of different types of data that
$\Omega_m=0.3$, the value $\Omega_{\Lambda}=0.7$ follows
immediately.

We now assert that Jones has {\it proved} his contention that
$\Omega_{\Lambda} =0.7$ implies a mass scale of $\sim 5 \ Tev/c^2$
(($5.8\times 10^{16} \ ^oK$) or visa versa. To our knowledge, this
is the first time that the Planck mass has been directly and {\it
quantitatively} connected to any observable phenomenon. We
emphasize that although the Jones argument connects a conventional
FRW universe to a denser situation where conventional concepts
loose operational meaning, {\it all} that he requires is that
normal scaling holds across the phase transition, and depends on
only one scale factor. That there is only one independent scale
factor is the conclusion acceptance of Occam's razor would
establish directly.

We also wish to emphasize that our reproduction of E.D.Jones
calculation here makes no claim to novelty, and is presented prior
to the posting of his own paper because of an unexpected delay in
the presentation of his own way of looking at the problem. We
stress that his thinking contains novel elements not discussed
here and that our cruder discussion should not be used as a
substitute for his work as soon as that becomes available.

\begin{center}
{\bf IV. Relation to Bit-String Physics}\\
\end{center}

We have seen that Jones' calculation depends on one fundamental
dimensionless scaling parameter, the number of Plancktons
($N_{Pk}=Z^2 \sim 4.3\times 10^{61}$) which thermalize at mass
scale $m \sim 5 \ Tev$ leaving behind energy density $\epsilon$
per Planckton. Since $m^2=\epsilon M_P$ (Eq. 12), any fundamental,
dimensionless theory which allows us to (a) identify within its
structure the Planck mass and three other dimensionless structural
constants which bear a known, and  mutually independent,
connection to, e.g. $c$, $\hbar$ and $k$ and (b) calculate $Z$ (or
$m$ or $\epsilon$) would allow us to say we have a first order,
fundamental understanding of physical cosmology. Where could we
find or how could we construct such a theory?

The candidate theory we examine here is bit-string
physics\cite{Noyes01}. We choose this construction both because
Jones' theory is historically connected to the research program
that led to bit-string physics (specifically, by his use of the
Dyson-Noyes argument), and because the only places where his
prediction that $\Omega_{\Lambda} \sim 0.6$ appear are in papers
stimulated by his private communication of that result to
HPN\cite{Noyes00a,Noyes00b}. Bit-string physics in turn arose out
of the combinatorial hierarchy of Amson, Bastin, Kilmister and
Parker-Rhodes\cite{Parker-Rhodes&Amson98}, which in turn came out
Bastin and Kilmister's\cite{Bastin&Kilmister95} interest in
Eddington's search for a fundamental theory\cite{Kilmister94}. One
reason this research program looks promising is, among other
things, because of Eddington's contention that dimensionless
numbers like $\alpha_{e^2} =\hbar c/e^2 \approx 137$ get into
physics only when we start from a logical or mathematical
``pre-physics'' and/or ``pre-geometry" construction.  Then such
dimensionless numbers could arise {\it prior} to development of
procedures we can relate to conventional measurement and which
give quantitative meaning to dimensional symbols like $\hbar, c$
and $e^2$ in any consistent system of physical units. To one of us
(HPN), the program universe construction which was created as part
of the research program on the combinatorial hierarchy sounds
suspiciously like Jones' postulated ``non-adiabatic process" which
takes the universe from the Planck scale up to a scale where
current physics clearly applies.

Unfortunately, none of this fundamental work on the combinatorial
hierarchy or bit-string physics has as yet led to a {\it reliable,
quantitative} connection to mainstream physics. Critically viewed,
one might say that this work consists of a few numerical
coincidences which are still in search of what might deserve to be
called a speculative physical theory. We contrast this with Jones'
theory, which we hope the first three sections have shown {\it
does} constitute a theory based on a few fundamental and generally
accepted principles in mainstream physics. Despite these critical
remarks, we include a brief discussion of the connection to
bit-string physics here in the hope that this discussion may help
in planning future research.

We start with the best succinct, accurate and published statement
of what the {\it combinatorial hierarchy} is. This is due to
McGoveran\cite{McGoveran88a}:

\begin{quote}
The Combinatorial Hierarchy is generated from two recursively
generated sequences. The first is governed by the recursion
formula $n_{i+1}= 2^{n_i}-1$ (a formula familiar to those who have
studied the Mersenne primes), and begins with the term n=2 leading
to the sequence $3, 7, 127, 2^{127}-1,....$ The cumulative
cardinals of this series (ignoring the initial term) also form a
series which has interpretive significance, namely $2, 3, 10, 137,
\sim 1.7016...\times 10^{38} + 137,...$

The second recursively generated sequence is governed by the
formula $m_{i+1} = m_i^2$. These two sequences have various
justifications. Perhaps the clearest presentation has been given
by Clive Kilmister (correspondence to H.P.Noyes date Oct. 16,
1978), paraphrased here as follows:

Definition:

By a combinatorial hierarchy is meant a collection of levels
selected as follows:

a) the elements of level L are a basis of a vector space $V/Z_2$

b) the elements at Level L+1 are non-singular (i.e. invertible)
linear operators mapping $V/Z_2$ into $V/Z_2$

c) each element A at level L+1 are mapped to a subset S of the
elements at Level L by the correspondence: the proper eigenvalues
of A [i.e, Av = v] are exactly the linear subspace generated by S.

d) each element at level L+1 is chosen independent, allowing the
process to be repeated for level L+2, L+3, L+4, ...

Theorem 1:

There is a unique hierarchy (up to isomorphism) with more than
3-levels and it has the following successive numbers of elements:
$2, 3, 7, 127, 2^{127} - 1$ and terminates at level 4 due to the
fact that the operators have $m^2$ elements if the vectors are
m-fold and $2^n$ (required for $V/Z_2$) increases too fast.

\end{quote}

What strikes some of us when we encounter the cumulative sequence
is that the third term $137 \approx \hbar c/e^2 = \alpha_{e^2}$ is
the inverse fine structure while the number of elements in the
terminating level $2^{127}+136 \approx 1.7016\times 10^{38}
\approx \hbar c/G_Nm_{proton}^2 = M_P^2/m_{proton}^2$ is the
square of the ratio of the Planck mass to the proton mass. Note
that one is the Dyson-Noyes number characterizing electromagnetic
interactions, and the other the corresponding Dyson-Noyes number
for the gravitating particles which constitute most of the known
paticulate mass of the universe (if the ``dark matter" can be
shown {\it not} to be particulate, or about 10\% of the total mass
of the universe if the ``dark matter" {\it is} particulate). This
looks promising for a cosmology based on this construction. But
progress toward a physical theory has been distressingly slow. A
historical summary has been published\cite{Noyes97}. Perhaps the
most encouraging results are the derivation of the Sommerfeld
formula and calculation of the next four significant figures
(beyond 137) in the inverse fine structure constant by
McGoveran\cite{McGoveran88b,McGoveran&Noyes91} and his various
combinatorial corrections to other results, reprinted and
discussed in\cite{McGoveran00}. All of these corrections improve
the fit to experiment.

Despite the vagueness of the contact between bit-string physics
and relativistic quantum particle dynamics, the structure already
discussed suggests a way to calculate two cosmological
parameters\cite{Noyes00a}. The first is the dark matter to
ordinary matter ratio. In the hierarchy construction, we do not
encounter the connection to electromagnetism until we have
constructed level 3 (i.e. the level characterized by $137= 127+10$
cumulative elements). However, one string in level 4 interacts
gravitationally with everything. This strongly suggests that that
the first two levels $10=3+7$ do not interact electromagnetically
but do interact gravitationally, and could therefore be used to
represent dark matter, whether it is particulate or
non-particulate. Then, if we use a constructive algorithm with an
arbitrary, stochastic element (as is done, for example, in {\it
program universe} --- see\cite{Noyes97} for details), the usual
assumption that in the absence of further information all elements
receive equal weight immediately predicts that the dark matter to
electromagnetically interacting matter ratio is $127/10 = 12.7$.
At the time of nucleosynthesis, it is also plausible to assume
that the the dark matter to electromagnetically interacting (at
the fundamental level) matter, can be approximated by the dark
matter to baryonic matter ratio, so we make that further
assumption. The final step needed to connect to cosmological
observation (see, e.g.\cite{PDG00,PDG02}) is the photon number to
baryon number ratio at the time of nucleosynthesis.

To get the photon/baryon ratio, we need only slightly more detail
about our constructive algorithm than has already been invoked. As
is discussed in more detail in the paper already
cited\cite{Noyes00a}, the most likely bit-strings a stochastic
construction yields, i.e. those (when of even length $2N$) with
the number of zeros ($N_0$) equal to the number of ones ($N_1$)
--- hence $N_0=N_1$, $N_0+N_1=2N$, are also the prime candidates
for representing photon labels (quantum numbers). The next most
likely strings have $N_0=N_1\pm 1$ or, with equal probability
$N_0=N_1\mp 1$. Note that these strings with an odd number of
ones, conserve this characteristic when ``interacting" with the
photon (even number of ones) strings as modeled by our basic
operation of addition in $Z_2$. This in turn suggests a conserved
quantum number such as baryon number. Further, the more detailed
interpretation of the basic {\it progam universe} algorithm in the
reference already cited\cite{Noyes97} seems to yield the type of
driving terms (two-body scatterings) needed for a finite particle
number relativistic quantum mechanical scattering
theory\cite{AKLN}.

The basic processes needed to estimate the photon baryon ratio in
the cosmological context of nucleosysnthesis we are discussing are
then the probability of a photon-photon scattering process with a
similar process as ``spectator" (in the few body sense) compared
to a photon-baryon scattering process with the same spectator. In
the first case there are four photons in the initial (and final)
state, and in the second there is one baryon and three photons in
the initial (and final) state. At this stage in the construction
(level 4 completed) the labels are strings of length 256. To
change the photon label to a baryon label in one of the four
photon labels, obtaining the next most probable process, can only
happen in one out of $256^4$ ways, giving a baryon-photon ratio of
$1/256^4 \sim 2.3\times 10^{-10}$. When first
presented\cite{Noyes00a}, this result was in comfortable agreement
with what was then known about this number from cosmic abundances
of the primordial nuclei, and predicted a value of $\Omega_m$
which was, perhaps, a little low, but amazingly good for such a
speculative calculation.

Recent observations, particulary of the primordial deuterium to
hydrogen ratio as inferred from the absorption spectra when the
interstellar and intergalactic deuterium is illuminated by very
early quasars, have tightened the allowed values of the
baryon-photon ratio and moved the median up nearly a factor of
two, as estimated by Fields and Sarkar\cite{Fields and Sarkar02}.
Their recommended limits now just exclude HPN's value and suggest
that there may have been an error in his reasoning. The error was
that the $1/256^4$ value ignored the fact that {\it both} the case
when $N_1$ exceeds $N_0$ by one {\it and} the case when it is one
less should have been counted, providing the ``missing" factor of
two. If accepted, this correction has the added advantage of
moving the prediction of $\Omega_m$ closer to the median of the
observed value. See\cite{Noyes00a} for the details of the original
calculation. We conclude here that HPN's calculation does provide
a weak indication that bit-strig physics predicts $\Omega_m \sim
0.3$, but that more work is badly needed before much confidence
can be placed in it.

If one accepts the bit-string prediction that $\Omega_m \sim 0.3$
{\it and} the empirical fact\cite{Frieman02b} that space is flat
to about 6\%, we can then use the constraint for flat space
$\Omega_{\Lambda} + \Omega_m =1$ to say that bit string physics
{\it predicts} $\Omega_{\Lambda} \sim 0.7$. Then, as we have seen
in our discussion in the first three sessions Jones'
MICROCOSMOLOGY shows that the dark energy density at the current
time can be understood, that thermalization/baryon number
conservation/physical space-time all became meaningful at a mass
scale $m \sim 5 Tev$, and all started from a pre-physics,
pre-geometry universe  characterized by a virtual energy of
$N_{Pk} \sim 4\times 10^{61}$ Planck masses.

Clearly this puts a high premium on tightening up the connection
between bit-string physics and mainstream physics to the point
where it can enter the field as a respectable contender for the
research interest of the physics and astrophysics community. Where
to begin is a matter of taste. One possibility, which HPN favors,
is to try to show that dark matter is particulate with a unique
mass of $m_D \sim 5 \ Tev/c^2$ and {\it only} gravitational
interactions. Searches for dark matter in this range are already
under way; having a prediction to confirm or refute might help the
experimenters. On the theoretical side, we note that bit-string
physics can already claim to have provided a kind of
strong-electromagnetic unification by arguing that the pion mass
can be approximately calculated as twice the inverse fine
structure constant in units of the electron mass. To get
weak-gravitational unification, we might argue by analogy with the
weak-electromagnetic unification (including the Coulomb force)
which gives the heavy vector mesons (W,Z) and one or more massive
Higgses, that electromagnetic-gravitation unification (including
Newton's gravitational force) might give a massive scalar at $\sim
5 \ Tev/c^2$. This is, admitttedly, a long shot, but might provide
amusement for those with appropriate skills.

\begin{center}
{\bf V. Conclusions}
\end{center}

Accepting the growing consensus that current observations
inescapably require that $\Omega_{\Lambda}= 0.7$ to $20\% $ or
better, Jones' reasoning allows him to use this value to state
that the mass-energy-temperature scale at which ``pre-physics''
makes a transition to observable space, time and particles is
$\sim 5 \ Tev$! In other words, the next generation of particle
accelerators will either substantiate the view that we {\it
already} have in hand enough physics to understand the basic
structure parameters of the currently observed universe {\it
quantitatively}, or the next generation of high energy particle
physics machines (if constructed) will necessarily provide graphic
evidence that fundamental revision of our basic concepts is
needed. We trust we have made it clear that bit-string physics has
the basic logical structure needed to provide the one theoretical
scaling factor on which Jones' theory rests. Optimists will go
further and think that it already provides weak support for
MICROCOSMOLOGY.

\begin{center}
{\bf Acknowledgment}
\end{center}

We are most grateful to E.D.Jones for permitting us to post this
discussion of his theory before his own paper is readily
available. We emphasize again that our paper should not be used as
a substitute for his original work, particularly since we have
used a number of short cuts and simplifications in our
presentation, and tied it to bit-string physics in a way that has
not been discussed with him.


\begin{thebibliography}{99}
\bibitem{Frieman02a}
Frieman\cite{Frieman02b} has noted that General Relativity {\it
requires} that the equation of state for the ``dark energy"
currently observed have a negative pressure.
\end{thebibliography}
\end{document}